\documentclass[aps,rmp,reprint,amsmath,amssymb,graphicx,longbibliography]{revtex4-1}

\usepackage{bm}
\usepackage{longtable}

\usepackage{amsmath,amscd,amsfonts,amssymb, amsbsy, color,bbm, bm} 
\usepackage{graphicx}
\usepackage{enumitem}
\newcommand{\be}{\begin{equation}}
\newcommand{\ee}{\end{equation}}

\newcommand{\ba}[1]{\left(\begin{array}{#1}}
\newcommand{\ea}{\end{array}\right)}
\newcommand{\alx}{\mbox{$\alpha_{1}$}}
\newcommand{\aly}{\mbox{$\alpha_{2}$}}
\newcommand{\alz}{\mbox{$\alpha_{3}$}}

\begin{document}

\title{Lorentz canoncial forms of two-qubit states}  

\author{Sudha} 
\affiliation{Department of Physics, Kuvempu University, 
	Shankaraghatta-577 451, Karnataka, India}
\affiliation{Inspire Institute Inc., Alexandria, Virginia, 22303, USA.}
\email{tthdrs@gmail.com}

\author{A. R. Usha Devi} 
\affiliation{Department of Physics, Bangalore University, 
	Bangalore-560 056, India}
\affiliation{Inspire Institute Inc., Alexandria, Virginia, 22303, USA.}
\email{ushadevi@bub.ernet.in}

\author{B. N. Karthik}
\affiliation{Department of Physics, Bangalore University, Bangalore-560 056, India}
\email{karthikbnj@gmail.com } 

\author{H. S. Karthik} 
\affiliation{International Centre for Theory of Quantum Technologies, University of Gd{\'a}nsk, Gd{\'a}nsk, Poland}
\email{karthik.hs@ug.edu.pl}

\author{Akshata Shenoy H} 
\affiliation{International Centre for Theory of Quantum Technologies, University of Gd{\'a}nsk, Gd{\'a}nsk, Poland}
\email{akshata.shenoy@ug.edu.pl}

\author{K. S. Mallesh}
\affiliation{Regional Institute of Education (NCERT), Mysuru 570006, India}
\email{ksmallesh@gmail.com}

\author{A. V. Gopala Rao}
\affiliation{Department of Studies in Physics,  University of Mysore, Manasagangotri, Mysore-570 006, India}
\email{garakali@gmail.com}

\date{\today{}}

\begin{abstract}
	The Bloch sphere provides an elegant way of visualizing  a qubit. Analogous representation of  the simplest
	composite state of two-qubits has attracted significant attention.  Here we present a detailed mathematical analysis of  the real-matrix parametrization and associated geometric picturization of  arbitrary two-qubit states - up to their  local SL(2C) equivalence -- in terms of canonical ellipsoids inscribed within the Bloch sphere.       
\end{abstract}


\maketitle
\tableofcontents{}

\section{Introduction}
\label{intro}
Bloch sphere~\cite{Bloch1946}, named after the physicist Felix Bloch,  provides a geometrical embedding for a qubit (quantum two-level system). Pure states of a qubit constitute the entire Bloch sphere and mixed states lie inside.  This picture has been utilized widely in quantum dynamics in general and  quantum information processing in particular. Extensions of similar  picturization for  higher-dimensional state spaces have been proposed~\cite{Kimura2005,zyzcBook2006, Sandeep2016,Rau}. These studies point towards the emergence of intricate geometric structures as the Hilbert space dimension goes up and this hinders their utility in the field of quantum information processing. Given the key role played by entanglement in quantum information processing, there have been dedicated efforts to unravel geometric features associated with the simplest bipartite system  viz., two-qubit state~\cite{hor96, verstraete2001,verstraete2002,avron2007,gamel2016,jevtic2014,MilneNJP2014,Rau,supra}. Visualization of two-qubit states as ellipsoids inscribed inside the Bloch ball is found to be useful to understand  quantum correlation features~\cite{verstraete2002,jevtic2014,MilneNJP2014,supra}. Investigation of the geometry of two-qubit states depends crucially on the local invertible linear transformations. In particular, restricting to local invertible qubit transformations  represented by $2\times 2$ complex matrices with determinant 1, enables one to exploit the homomorphism between the groups SL(2,C) and Lorentz group SO(3,1). In this paper we address the intricate and subtle features of Lorentz canonical forms associated with two-qubit states, which lead to their geometrical visualization as ellipsoids inside the Bloch ball.  

We organize our paper in the following manner: Section II is devoted to the discussion of $4\times 4$ real matrix parametrization $\Lambda$ of two-qubit density matrix $\rho_{AB}$ and its transformation properties when individual qubits are subjected to SL(2,C) operations. Introduction of a real symmetric $4\times 4$ matrix $\Omega=\Lambda^T\,G\,\Lambda$, where $G={\rm diag}\,(1,-1,-1,-1)$ and the spectral analysis of the matrix $G\,\Omega$ constitutes the content of Section III.  Based on the analysis of the eigenvalues and eigenvectors of $G\,\Omega$ we arrive at {\em two} types of  the Lorentz canonical forms of $\Lambda$ and the associated   canonical forms of the  two-qubit density matrices  sharing the same SL(2,C) orbit in Section IV, which also contains details on the geometrical embedding of the two-qubit states as {\em canonical steering ellipsoids} inside the Bloch sphere.  Section V has illustrative physical examples capturing the spectral analysis of $G\Omega$ and the geometrical visualization of the two-qubit density matrix $\rho_{AB}$.  We summarize our results in Section VI.    
 
\section{Real parametrization of two-qubit density matrix and Lorentz transformations} 
\label{sec2}
Let us denote the Pauli basis 
\begin{equation}
\sigma_\mu\equiv\{\sigma_0=\mathbbm{1}_2,\sigma_0=\sigma_x,\sigma_2=\sigma_y,\sigma_3=\sigma_z\}
\end{equation} where $\mathbbm{1}_2$, denotes $2\times 2$ identity matrix and $\sigma_x,\sigma_y,\sigma_z$ are the Pauli spin matrices. Any arbitrary two-qubit state $\rho_{AB}$ can be expressed in the Pauli tensor basis $\{\sigma_\mu\otimes \sigma_\nu, \mu,\nu=0,1,2,3\}$ as,
\begin{eqnarray}
\label{rho}
\rho_{AB}&=&\frac{1}{4}\, \sum_{\mu,\,\nu=0}^{3}\,   
\Lambda_{\mu \, \nu}\, \left( \sigma_\mu\otimes\sigma_\nu \right)  
\end{eqnarray}
where 
\begin{eqnarray}
\label{lambda}
\Lambda_{\mu \, \nu}&=& {\rm Tr}\,\left[\rho_{AB}\,
(\sigma_\mu\otimes\sigma_\nu)\,\right]. 
\end{eqnarray} 
Hermiticity of the density matrix implies that $\Lambda_{\mu \, \nu}$ are real for all $\mu,\nu$.  
 
Expressed in the $2\times 2$ block form, the $4\times 4$ real matrix $\Lambda$ (see   (\ref{lambda})) assumes the following compact form,   
\begin{eqnarray}
\label{block}
\Lambda&=&\left(\begin{array}{ll} 1& {\bf b}^T  \\ {\bf a}  &  T   
\end{array}\right).
\end{eqnarray} 
Here the superscript ``$T\,$'' denotes matrix transposition; ${\bf a}=(a_1,\, a_2,\, a_3)^T$, ${\bf b}=(b_1,\, b_2,\ b_3)^T$ denote  Bloch vectors  of the reduced density matrices  of qubits $A$, $B$:
\begin{eqnarray}
\rho_A&=&{\rm Tr}_B(\rho_{AB}) \nonumber \\ 
&=& \frac{1}{2}\, \sum_{\mu=0}^{3}\,   
a_{\mu}\,  \sigma_\mu,\ \  a_{\mu}=(1,\,{\bf a}) \\ 
 \rho_B&=&{\rm Tr}_A(\rho_{AB}) \nonumber \\ 
 &=& \frac{1}{2}\, \sum_{\mu=0}^{3}\,   
 b_{\mu}\,  \sigma_\mu,\ \  b_{\mu}=(1,\,{\bf b}) \\ 
 \end{eqnarray}
and $T$ denotes the $3\times 3$ real correlation matrix~\cite{hor96}, elements of which are given by $t_{ij}={\rm Tr}(\rho_{AB}\,\sigma_i\otimes\sigma_j),\, i,j=1,2,3$. The $4\times 4$  real matrix $\Lambda$ is thus characterized by 15 real parameters (3 each of the   Bloch vectors ${\bf a},\ {\bf b}$ and 9 elements of the correlation matrix $T$) and it provides a faithful {\em real matrix parametrization} of the two-qubit density matrix  $\rho_{AB}$.


\subsection{Lorentz transformation on the real parametrization matrix ${\bm\Lambda}$ of two-qubit density matrices}
\label{2a}
Under the action of local invertible operations  $A, B\in {\rm SL(2,C)}$ on individual qubits the two-qubit density operator transforms as~\cite{verstraete2001, jevtic2014, supra}  
\be
\label{slocc2}
\rho_{AB}\longrightarrow\, \widetilde{\rho}_{AB} =\frac{\left(A\otimes\, B\right)\, \rho_{AB}\, \left(A^\dag\otimes B^\dag\right)}{{\rm Tr}\left[\,\rho_{AB}\,(\, A^\dag A\otimes B^\dag B\, )\,\right]},
\ee
 In view of the  homomorphism between the groups  ${\rm SL(2,C)}$ -- elements of which are  $2\times 2$ complex invertible matrices with determinant 1 --  and the orthochronous proper Lorentz group 
 SO(3,1) -- consisting of $4\times 4$ real matrices $L$ which preserve the Minkowski metric $G={\rm diag}\,(1,-1,-1,-1,-1)$ such that  $L^T\, G\, L=G$ and $\det\,L=1$ --   there exists a correspondence~\cite{KNS}  $\pm A \mapsto L_A,\, \pm B \mapsto L_B $, for $A,\,B\in$ SL(2,C)  and $L_A,\, L_B\in$ SO(3,1) .  More specifically, the Pauli basis matrices    $\sigma_\mu\otimes\sigma_\nu, \mu,\, \nu=0,1,2,3$  transform  under SL(2,C)\, $\otimes$\,SL(2,C) as  
\begin{eqnarray}
\label{oplt}
&&(A\otimes B)(\sigma_\mu\otimes\sigma_\nu)(A^\dag\otimes B^\dag) = 
A\,\sigma_\mu A^\dag
\otimes B\,\sigma_\nu B^\dag    \nonumber \\
&& \hskip 0.5in =  \sum_{\alpha,\beta=0,1,2,3} \left(L_{A}\right)_{\alpha\mu} \left(L_{B}\right)_{\beta\nu} \sigma_\alpha\otimes\sigma_\beta.  
\end{eqnarray} 
Thus, the transformation  $\rho_{AB}~\rightarrow~ \tilde{\rho}_{AB}$ on the two-qubit state is equivalent to the Lorentz transformation (up to normalization) of the real matrix $\Lambda$: 
\begin{eqnarray}
\label{sl2c}
\Lambda\longrightarrow \tilde{\Lambda}&=&\, L_A\,\Lambda\, L_B^{T}.
\end{eqnarray}  
 Obviously  $\tilde{\Lambda}$, obtained after the Lorentz transformations $L_A, L_B$  on $\Lambda$  (see (\ref{sl2c})),   parametrizes  the two-qubit density matrix $\tilde{\rho}_{AB}=\frac{\left(A\otimes\, B\right)\, \rho_{AB}\, \left(A^\dag\,\otimes B^\dag\right)}{{\rm Tr}\left[\,\rho_{AB}\,(\, A^\dag A\otimes B^\dag B\, )\,\right]}.$  
 
\subsection{Real symmetric matrix $\mathbf{\Omega=\Lambda^T\,G\,\Lambda}$} 
\label{2b}
Let us denote the set  of all non-negative operators acting on the qubit Hilbert space $\mathbbm{C}_2$ by ${\mathcal{P}}^+ :=\{P \vert P\, \geq\, 0\, \}$. Any element  $P\in {\mathcal{P}}^+$ can be 
expressed  in the Pauli basis $\sigma_\mu=(\mathbbm{1}_2,\sigma_1,\sigma_2,\sigma_3)$ as 
\begin{equation}
\label{pmu}
P=\frac{1}{2}\,\sum_{\mu}\, p_\mu\,  \sigma_\mu
\end{equation}   
where $$p_\mu= {\rm Tr}(P\, \sigma_\mu), \mu=0,1,2,3$$ are the four real parameters characterizing $P$. With every $P\in {\mathcal{P}}^+$, one can associate  a Minkowski four-vector~\cite{KNS,synge}  $$\mathbf{p}=(p_0,\, p_1,\, p_2,\, p_3)^T.$$ Non-negativity  $P\geq 0$  is synonymous to the conditions  ${\rm Tr}\,(P)=p_0> 0$ and $\det P=p_0^2-p_1^2-p_2^2-p_3^2\geq 0$ on the four-vector $\mathbf{p}$ and thus $P$ can be recognized as an element of a positive-operator valued measure (POVM).  

In the language of  Minkowski space,  equipped with the metric 
$$G={\rm diag}\,(1,-1,-1,-1)$$ the positivity $P\geq 0$  is synonymous to the four-vector norm condition   $\mathbf{p}^T\, G\, \mathbf{p}\geq 0$,  along with the restriction $p_0> 0$ on zeroth component of the four-vector $\mathbf{p}$. In other words, the four-vector $\mathbf{p}=(p_0,\, p_1,\, p_2,\, p_3)^T$ of the POVM element $P$ is either {\em time-like} ($\mathbf{p}^T\, G\, \mathbf{p}> 0$) or {\em null} ($\mathbf{p}^T\, G\, \mathbf{p}=0$).    

Let us consider the map  
\begin{eqnarray}
\label{mapA2B}
P \mapsto Q&=& 
2\, {\rm Tr}_B \left[\rho_{AB}\, (\mathbbm{1}_2\otimes P)\right]
\end{eqnarray}
from  the set of all non-negative operators ${\mathcal{P}}^+~:=~ \{P\, \vert\ P\geq 0 \}$ on the Hilbert space ${\mathcal{H}}_B$  to the set of non-negative operators ${\mathcal{Q}}^+:=\, \{Q\, =2\,  {\rm Tr}_B[\rho_{AB}\,( \mathbbm{1}_2\otimes P)]\}$. This map has the following physical interpretation:  When Bob carries out a measurement on his qubit and obtains outcome linked with the POVM element $P$, Alice's qubit gets steered to the quantum state proportional to  $Q={\rm Tr}_B \left[\rho_{AB}\, (\mathbbm{1}_2\otimes P)\right]$.    

It is readily seen that 
\begin{eqnarray}
\label{paqbdef}
Q&=&2\,{\rm Tr}_B\left[\rho_{AB}\, (\mathbbm{1}_2\otimes P)\right]\nonumber \\ 
&=&\frac{1}{2}\, \sum_{\nu}\, \left( \Lambda\,\mathbf{p} \right)_{\nu}\,  \, \sigma_{\nu}
\end{eqnarray} 
 resulting in the Minkowski four-vector transformation 
\begin{equation} 
\label{pos1}
\mathbf{q}=\Lambda\,\mathbf{p}.
\end{equation} 
In other words, the map $P \mapsto Q$ is identical to the four-vector map  $\Lambda: \mathbf{p}\mapsto \mathbf{q}=\Lambda\,\mathbf{p}$.

Clearly, the squared Minkowski norm of the four-vector  $\mathbf{q}$ is non-negative (i.e., the four-vector $\mathbf{q}$ is either time-like or null). Thus,        
\begin{eqnarray}
\label{pos2}
\mathbf{q}^T\, G\, \mathbf{q}\geq 0 &&\Longrightarrow  \ 
\mathbf{p}^T\,\Lambda^T\, G\, \Lambda\, \mathbf{p}\geq 0  \nonumber \\ 
&&\Longrightarrow \  \mathbf{p}^T\,\Omega\, \mathbf{p}\geq 0
\end{eqnarray}
 where 
\be
\label{oA}
\Omega=\Lambda^T\, G\, \Lambda
\ee
denotes a real symmetric $4\times 4$ matrix, associated with the real parametrization $\Lambda$ of the two-qubit density matrix $\rho_{AB}.$  
Furthermore, positivity of the zeroth component of the four-vector $\mathbf{p}$ (i.e., ${\rm Tr}\, (P)>0$) imposes that  
\begin{equation}
\label{pa0pos}
p_{0}>0 \, \Longrightarrow q_{0}=\left(\Lambda\,\mathbf{p}\right)_0> 0.
\end{equation}

The $4\times 4$ real symmetric matrix $\Omega=\Lambda^T\, G\, \Lambda$ constructed from the real counterpart $\Lambda$ of the two-qubit density matrix $\rho_{AB}$ plays a key role in our analysis and is inspired by the methods developed in classical polarization optics~\cite{AVG1,AVG2}.  Spectral analysis of the $4\times 4$ matrix $G\,\Omega=G\,\Lambda^T\, G\, \Lambda$ results in the geometrical visualization of two-qubit states on the SL(2,C) orbit as ellipsoids inside the Bloch Ball.

\section{Spectral analysis of the matrix ${\mathbf G}\,{\bm\Omega}$ }
\label{sec3}
We focus on the  condition (see equations (\ref{pos2}), (\ref{oA}) and (\ref{pa0pos}))
\be
\label{a2}
\left\{{\mathbf q}=\Lambda\,{\mathbf p}\vert {\mathbf q}^T\,G{\mathbf q}={\mathbf p}^T\,\Omega\,{\mathbf p}\geq 0, \ \ q_0> 0\right\}
\ee

Since every time-like four-vector  can always be expressed as a sum of null four-vectors~\cite{KNS}, we choose to confine ourselves to the set of null four-vectors 
\be
\label{a3}
\left\{{\mathbf p}\equiv{\mathbf p}_n=(1,{\mathbf x})^T, \ \  {\mathbf x}^T{\mathbf x}=1; \ \ 
{\mathbf p}_n^T\,G\,{\mathbf p}_n=0\right\}
\ee
without any loss of generality. 

The non-negativity constraint (\ref{a2}) can thus be 
expressed as 
\be
\label{a4}
\left\{{\mathbf q}=\Lambda\,{\mathbf p}_n\vert {\mathbf p}_n^T\,G{\mathbf p}_n=0\Longrightarrow {\mathbf q}^T\,G{\mathbf q}={\mathbf p}_n^T\,\Omega{\mathbf p}_n\geq 0\right\}
\ee
where $\Omega=\Lambda^T\,G\,\Lambda=\Omega^T$.

Under the transformation  (see equation (\ref{sl2c})) $\Lambda\rightarrow \widetilde{\Lambda}=L_A\,\Lambda\, L_B^T, \ L_A,L_B\in$\,SO(3,1), the matrix $\Omega=\Lambda^T\,G\,\Lambda$ gets subjected to a Lorentz congruent operation: 
\begin{eqnarray}
	\label{oBc}
		\widetilde{\Omega}&=& {\widetilde{\Lambda}}^T\,G\,\widetilde{\Lambda} \nonumber \\ 
	&=&\left( L_{A}\,\Lambda\, L_{B}^{T}\right)^T\,G\,\left(L_{A}\,\Lambda\, L_{B}^{T}\right)\nonumber \\
	& =& L_{B}\, \Lambda^T\, L_{A}^T\, G \, L_{A}\, \Lambda\, L_{B}^T \nonumber \\ 
	& =& L_{B}\, \Lambda^T\,  G  \Lambda\, L_{B}^T \nonumber \\
	&=& L_{B}\, \Omega\, L_{B}^T
\end{eqnarray}
 where we have used the metric-preserving property $L_{A}^T\, G \, L_{A}=G$ of  $L_A\in$ SO(3,1).

With the help of the  relation 
$$G\, L\,=\left(L^T\right)^{-1}\, G$$ 
satisfied by  Lorentz transformation matrix, it is easy to identify that    
\begin{eqnarray}
\label{goab}
 G\, \Omega= G\,\Lambda^T\, G\, \Lambda  
\end{eqnarray}
undergoes a similarity transformation as follows:     
\begin{eqnarray}
\label{gobs}
G\, \Omega\, &\longrightarrow&\, G\,  L\,  \Omega \, L^T  \nonumber  \\
&=& \left(L^T\,\right)^{-1}\, G\,\Omega\, L^T. 
\end{eqnarray}
It is thus clear that {\em the eigenvalues of the $4\times 4$ real matrix $G\, \Omega$ remain invariant under Lorentz transformation}. We refer to the eigenvalues and eigenvectors of $G\, \Omega$ as $G$-eigenvalues and $G$-eigenvectors of the matrix $\Omega$ respectively. Detailed discussion on the spectral properties of the matrix $G\,\Omega$ associated with two-qubit states is presented in the following subsection.  

\subsection{${\bm G}$-eigenvalues  of  $\bm{\Omega}$}
\label{3a}

Let us express the matrix $\Omega$ in the block form
\be
\label{oblock}
\Omega=\ba{cc} \Omega_{00} & {\bm \omega}^T \\ {\bm \omega} & S  \ea
\ee 
where  ${\bm \omega}^T=(\Omega_{01},\,\Omega_{02},\,\Omega_{03})$ and $S=S^T$ is a real, symmetric $3\times 3$ matrix.  We consider a simpler form of the matrix $\Omega$ with the help of Lorentz congruent transformations $\widetilde{\Omega}=L^T\,\Omega\,L$ (see equation (\ref{oBc})). To this end, we choose $L=1\oplus R$, where $R\in$ SO(3) is a $3\times 3$ rotation matrix diagonalizing the real symmetric matrix $S$ (see equation (\ref{oblock}) i.e.,  $R\,S\,R^T=S_\alpha={\rm diag}\,(\alpha_1,\alpha_2,\alpha_3)$. Consequently, ${\bm \omega}\rightarrow R\,{\bm \omega}$. We thus obtain

\begin{eqnarray}
\label{ac}
\Omega\rightarrow \Omega_0= L^T\,\Omega\,L&=& (1\oplus R)\, \Omega\, (1\oplus R^T)=\ba{cc} n_0 & {\mathbf n}^T \\ {\mathbf n} & S_\alpha  \ea \nonumber \\
\end{eqnarray}
where we have denoted $\Omega_{00}=n_0$ and $R\,{\bm \omega}={\mathbf n}=(n_1,\,n_2,\,n_3)^T$. 

We express the positivity condition $\mathbf{p}_n^T\,\Omega_0\,{\mathbf{p}_n}\geq 0$  (see equation (\ref{a4}), where we consider the simpler structure $\Omega_0$ given by equation (\ref{ac}):
\begin{eqnarray} 
\label{ds1}
D(\Omega_0;\,{\mathbf x})&=&n_0+2\,{\mathbf x}^T\,{\mathbf n}+{\mathbf x}^T S_\alpha\, {\mathbf x}\geq 0  \\ && \hskip 0.5in  \ \ \forall \  {\mathbf x}^T\,{\mathbf x}=1.\nonumber
\end{eqnarray}
For condition (\ref{ds1}) to hold good, the absolute minima of the function $D(\Omega_0;\,{\mathbf x})$, or equivalently,  {\emph {critical values}} of $D(\Omega_0;\,{\mathbf x})$ should be non-negative. In order to find the critical values of  $D(\Omega_0;\,{\mathbf x})$ we employ the Lagrange multiplier method and define 
\be
\label{k1}
K(\Omega_0;\,{\mathbf x})=D(\Omega_0;\,{\mathbf x}) + \lambda\, ({{\mathbf x}}^T{\mathbf x}-1)
\ee 
where $\lambda$ denotes the Lagrange multiplier.  

Setting the partial derivatives of $K(\Omega_0;\,{\mathbf x})$ with respect to $\lambda$ and $x_i,\, i=1,2,3$ equal to zero  at the critical values $\lambda_c,\, {\mathbf x}_{c}$ i.e., 
\begin{eqnarray}
\left. \frac{\partial K(\Omega_0;\,{\mathbf x})}{\partial \lambda}\right\vert_{\lambda_c,\, {\mathbf x}_{c}}&=&0\nonumber \\ 
\left.\frac{\partial K(\Omega_0;\,{\mathbf x})}{\partial x_i}\right\vert_{\lambda_c,\, {\mathbf x}_{c}}&=&0,\ \ i=1,2,3  
\end{eqnarray}
we obtain 
\be
\label{decoup}
(S_\alpha+\lambda_c\,\mathbbm{1}_3)\,{\mathbf x}_c=-{\mathbf n},\ \    {{\mathbf x}_c}^T\, {\mathbf x}_c=1 
\ee
or 
\be
\label{xnrel}
{x}_{ci}=\frac{-n_i}{(\alpha_i+\lambda_c)}, \ i=1,\,2,\,3. 
\ee
Substituting equation (\ref{xnrel})  and ${\mathbf x}_c^T \,{\mathbf x}_c= x_{c1}^2+x_{c2}^2+x_{c3}^2=1$  in equation (\ref{ds1}) we arrive at the constrained critical values $D_c$ of $D(\Omega_0;\,{\mathbf x})$:  
\be  
\label{d2}
D_c=n_0-\lambda_c-\sum_{i=1}^3\, \left(\frac{n_i^2}{\lambda_c+\alpha_i}\right).  
\ee 

Let us consider   
\be
\label{hfunction}
h(\lambda)=n_0-\lambda-\sum_{i=1}^3\, \frac{n_i^2}{\lambda+\alpha_i}, 
\ee 
obtained by replacing $\lambda_c$  in  $D_c$ (see equation (\ref{d2})) by a continuous variable $\lambda$. 

We list the following properties of the function $h(\lambda)$: 
\begin{enumerate}
\item From equation (\ref{hfunction}) it is seen that  $h(\lambda)$ has finite number of  discontinuities positioned at $\lambda=-\alpha_i$, $i=1,\,2,\,3$, whenever the corresponding $n_i\neq 0$.
\item From explicit evaluation we recognize that  
\be
h(\lambda)=-\frac{\mbox{det}\,(G\Omega_0-\lambda\,\mathbbm{1}_4)}{\mbox{det}\,(S_\alpha+\lambda\,\mathbbm{1}_3)}.  
\ee
This leads to an important observation that {\em the real roots of $h(\lambda)$ are nothing but the $G$-eigenvalues of the matrix $\Omega_0$}: 
\begin{eqnarray}
	\label{hfndet}
\mbox{det}(G\Omega_0-\lambda\,\mathbbm{1}_4)&=&-h(\lambda)\, {\mbox{det}\,(S_\alpha+\lambda\,\mathbbm{1}_3})  \\ 
&=&-h(\lambda)\, (\alpha_1+\lambda) (\alpha_2+\lambda) (\alpha_3+\lambda). \nonumber 
\end{eqnarray} 
\item The function $h(\lambda)$ changes sign across its discontinuity and is positive to the immediate left and negative to the immediate right of a discontinuity. This implies that  odd number of zeros (at least one) exist between any two discontinuities of $h(\lambda).$ 
\item From equation (\ref{hfunction}) it may be noticed that   $h(\lambda)\rightarrow \infty$ as $\lambda\rightarrow -\infty$ and  $h(\lambda)\rightarrow -\infty$ as $\lambda\rightarrow \infty$.  
Based on this observation, together with the behaviour of $h(\lambda)$ near a given discontinuity, it is inferred that even number of zeros (at least two) lie in 
the interval $(\alpha_{\rm max},\,\infty)$, where $\alpha_{\rm max}$ is the largest of $(\alpha_1,\alpha_2,\alpha_3).$
\item As $h(\lambda)\rightarrow -\infty$ as $\lambda\rightarrow \infty$ and the largest zero $\lambda_{\rm max}$ occurs in the interval $(\alpha_{max},\,\infty)$, the slope of $h(\lambda)$ at $\lambda_{\rm max}$ is either negative or zero.
\end{enumerate}

Depending on the number of non-zero values of $n_1,n_2,n_3$ and based on the degeneracies $\alpha_1$, $\alpha_2$, $\alpha_3$, there exist 20 possible situations, each with different number of discontinuities and zeros  of the function 
$h(\lambda)$: (i) none of $n_1,n_2,n_3$ are zero; (ii) one of $n_1,n_2,n_3$ is zero (i.e., $n_1=0$, $n_2,n_3\neq 0$; $n_2=0$, $n_1,n_3\neq0$; $n_3=0$, $n_1,n_2\neq 0$),
(iii) two of $n_1,n_2,n_3$ are zero (i.e., $n_1,n_2=0 \neq n_3$; $n_2,n_3=0\neq n_1$; $n_1,n_3=0\neq n_2$),
(iv) $n_1=n_2=n_3=0$. 

Each of these 4 cases fall under 5 different subclasses corresponding to the degeneracies of $\alpha_1,\alpha_2,\alpha_3$: non-degenerate i.e., (A)~$\alpha_1\neq\alpha_2\neq \alpha_3$, two-fold degenerate i.e,  (B1)~$\alpha_1=\alpha_2\equiv\alpha\neq \alpha_3$, (B2)~$\alpha_1\neq\alpha_2= \alpha_3\equiv\alpha$,
(B3)~$\alpha_1=\alpha_3\equiv\alpha\neq \alpha_2$, and fully degenerate i.e., (C)~$\alpha_1=\alpha_2=\alpha_3=\alpha$.

Associated with these $4\times 5=20$ distinct possibilities one may list the number of discontinuities and zeros of the function $h(\lambda)$. More specifically, it is seen that if $k\leq 3$ denote the finite number of distinct discontinuities of the function $h(\lambda)$, the function $h(\lambda)$ must have, at least $k+1$ real zeros.  When there are two distinct discontinuities
($k = 2$) at least {\em one} real zero of $h(\lambda)$ occurs
between them and  {\em two} zeros  (either distinct
or doubly repeated) appear in the region $(\alpha_{\rm max},\infty)$. Thus, 1 + 2 = 3 real zeros exist for $h(\lambda)$ when $k = 2$.  
In order to investigate the number $k$ of discontinuities and its association with the number of zeros of  $h(\lambda)$,  we express  the characteristic polynomial $\phi(\lambda)=\mbox{det}\left(G\Omega_0-\lambda\,\mathbbm{1}_4\right)$ of $G\,\Omega_0$ as 
\be
\phi(\lambda)= \phi_1(\lambda)\, g(\lambda)\,h(\lambda)
\ee
in terms of two simple poynomials $\phi_1(\lambda)$ and  $g(\lambda)$ such that (i) they have real roots; (ii) they are finite at every zero of the function $h(\lambda)$.  Explicit structure of     the polynomials $\phi_1(\lambda),$  $g(\lambda)$ and $h(\lambda)$ in each of the 20 different cases is given in Table~1. Based on careful  examination of the characteristic equation $\phi(\lambda) =\mbox{det}\left(G\Omega_0-\lambda\,\mathbbm{1}_4\right)=0$ of  $G\Omega_0$ and the explicit structures  of $\phi_1(\lambda)$, $g(\lambda)$, $h(\lambda)$ in each of the
20 cases listed in Table~1, we reach the following conclusion: If $r$ denotes the number of roots of $\phi_1(\lambda)$ and
	$k$ denotes the number of discontinuities of $h(\lambda)$, then  $r + k + 1 = 4$ in all 20 cases.  It is found that  whenever the number of zeros   $k+1$ of $h(\lambda)$ is less than 3 (or in turn, the number $k$ of discontinuities  exhibited by the function $h(\lambda)$ is less than 2),   then the roots of the polynomial $\phi_1(\lambda)$ determine the remaining $r$ number of $G$-eigenvalues of $\Omega_0$. This in turn culminates in the determination of all  {\em four}  $G$-eigenvalues $\lambda_\mu, \ \mu=0,1,2,3$ of the $4\times 4$ matrix $\Omega_0$.

\begin{widetext} 
\subsection{${\bm G}$-eigenvectors of $\bm{\Omega}$}
\label{sec4}

\begin{table}
			\caption{The polynomial functions $\phi_1(\lambda),\ g(\lambda)$ and $h(\lambda)$ in 20 different physical cases}
			\medskip
			\begin{tabular}{|c|c|c|c|c|}
				\hline
				Sl. No. & Case & $\phi_{1}(\lambda)$ & $g(\lambda)$ &
				$h(\lambda)$  \\
				\hline 
				(i)	& 	$n_1,\, n_2,\, n_3\neq 0$ & & &  \\ 
				\cline{1-2} 
				(A)	& $\alpha_1\neq\alpha_2\neq\alpha_3$ & 1 &$(\alx+\lambda)(\aly+\lambda)(\alz+\lambda)$ 
				&$n_0-\lambda-\sum_{i}\frac{n_i^2}{(\alpha_i+\lambda)}$  \\ 
				\hline 
				(B1) & 	$\alpha_1=\alpha_2=\alpha\neq \alpha_3$  &
				$(\alpha+\lambda)$&$(\alpha+\lambda)(\alz+\lambda)$
				&$n_0-\lambda-\frac{n_1^2+ n_2^2}{(\alpha+\lambda)}- \frac{n_3^2}{(\alpha_3+\lambda)}$  \\
				\hline
				(B2)	& $\alpha_1\neq \alpha_2=\alpha_3=\alpha$ &
				$(\alpha+\lambda)$&$(\alpha+\lambda)(\alpha_1+\lambda)$ &
				$n_0-\lambda- \frac{n_1^2}{(\alpha_1+\lambda)}-\frac{n_2^2+ n_3^2}{(\alpha+\lambda)}$  \\	
				\hline
				(B3)  	& $\alpha_1= \alpha_3=\alpha\neq \alpha_2$ &
				$(\alpha+\lambda)$&$(\alpha+\lambda)(\alpha_2+\lambda)$ &
				$n_0-\lambda- \frac{n_1^2+n_3^2}{(\alpha+\lambda)}-\frac{n_2^2}{(\alpha_2+\lambda)}$  \\	
				\hline
				(C)  	& $\alpha_1= \alpha_2=\alpha_3=\alpha$ &
				$(\alpha+\lambda)^2$&$(\alpha+\lambda)$ &
				$n_0-\lambda- \frac{n_1^2+n_2^2+n_3^2}{(\alpha+\lambda)}$  \\	
				\hline
				&&&&\\
				(ii)	& 	$n_1,n_2\neq 0$, $n_3 = 0$ & & &  \\ 
				\cline{1-2}
				
				(A)	& $\alpha_1\neq\alpha_2\neq\alpha_3$ & $(\alpha_3+\lambda)$ &$(\alpha_1+\lambda)(\alpha_2+\lambda)$ 
				&$n_0-\lambda-\frac{n_1^2}{(\alpha_1+\lambda)}+\frac{n_2^2}{(\alpha_2+\lambda)}$  \\ 
				\hline 
				(B1) & 	$\alpha_1=\alpha_2=\alpha\neq \alpha_3$  &
				$(\alpha_3+\lambda)(\alpha+\lambda)$&$(\alpha+\lambda)$
				&$n_0-\lambda-\frac{n_1^2+n_2^2}{(\alpha+\lambda)}$  \\
				\hline
				(B2)	& $\alpha_1\neq \alpha_2=\alpha_3=\alpha$ &
				$(\alpha+\lambda)$&$(\alpha_1+\lambda)(\alpha+\lambda)$ &
				$n_0-\lambda-\frac{n_1^2}{(\alpha_1+\lambda)}+ \frac{n_2^2}{(\alpha+\lambda)}$ 
				\\	
				\hline
				(B3)  	& $\alpha_1= \alpha_3=\alpha\neq \alpha_2$ &
				$(\alpha+\lambda)$&$(\alpha+\lambda)(\alpha_2+\lambda)$ &
				$n_0-\lambda- \frac{n_1^2}{(\alpha+\lambda)}-\frac{n_2^2}{(\alpha_2+\lambda)}$  \\	
				\hline 
				(C)  	& $\alpha_1= \alpha_2=\alpha_3=\alpha$ &
				$(\alpha+\lambda)^2$&$(\alpha+\lambda)$ &
				$n_0-\lambda- \frac{n_1^2+n_2^2}{(\alpha+\lambda)}$  \\	
				\hline
				(iii)	& 	$n_1=n_3=0$, $n_2\neq 0$ & &&\\
				\cline{1-2} 
				(A)	& $\alpha_1\neq\alpha_2\neq\alpha_3$ & $(\alpha_1+\lambda)(\alpha_3+\lambda)$ &$(\alpha_2+\lambda)$ 
				&$n_0-\lambda-\frac{n_2^2}{(\alpha_2+\lambda)}$  \\ 
				\hline 
				(B1) & 	$\alpha_1=\alpha_2=\alpha\neq \alpha_3$  &
				$(\alpha+\lambda)\,(\alpha_3+\lambda)$&$(\alpha+\lambda)$
				&$n_0-\lambda-\frac{n_2^2}{(\alpha+\lambda)}$  \\
				\hline
				(B2)	& $\alpha_1\neq \alpha_2=\alpha_3=\alpha$ &
				$(\alpha_1+\lambda)(\alpha+\lambda)$&$(\alpha+\lambda)$ &
				$n_0-\lambda-\frac{n_2^2}{(\alpha+\lambda)}$  \\	
				\hline
				(B3)  	& $\alpha_1= \alpha_3=\alpha\neq \alpha_2$ &
				$(\alpha+\lambda)^2$&$(\alpha_2+\lambda)$ &
				$n_0-\lambda- \frac{n_2^2}{(\alpha_2+\lambda)}$  \\	
				\hline 
				(C)  	& $\alpha_1= \alpha_2=\alpha_3=\alpha$ &
				$(\alpha+\lambda)^2$&$(\alpha+\lambda)$ &
				$n_0-\lambda- \frac{n_2^2}{(\alpha+\lambda)}$  \\	
				\hline
				(iv)	& 	$n_1=n_2=n_3=0$  & & &  \\ 
				\cline{1-2} 
				(A)	& $\alpha_1\neq\alpha_2\neq\alpha_3$ & $(\alpha_1+\lambda)(\alpha_2+\lambda)(\alz+\lambda)$ & $1$
				&$n_0-\lambda$  \\ 
				\hline 
				(B1) & 	$\alpha_1=\alpha_2=\alpha\neq \alpha_3$  & $(\alpha+\lambda)^2(\alpha_3+\lambda)$ &$1$ 
				&$n_0-\lambda$  \\ \hline
				(B2)	& $\alpha_1\neq \alpha_2=\alpha_3=\alpha$ & $(\alpha+\lambda)^2(\alpha_1+\lambda)$ &$1$ 
				&$n_0-\lambda$  \\ \hline
				(B3)  	& $\alpha_1= \alpha_3=\alpha\neq \alpha_2$ & $(\alpha+\lambda)^2(\alpha_2+\lambda)$ &$1$ 
				&$n_0-\lambda$  \\ \hline
				(C)  	& $\alpha_1= \alpha_2=\alpha_3=\alpha$ &
				$(\alpha+\lambda)^3$&$1$ &
				$n_0-\lambda$  \\	
				\hline					
			\end{tabular}
\end{table} 
\end{widetext}

Having identified that the $G$-eigenvalues of $\Omega_0$ are real roots of the function $h(\lambda)$, we proceed to investigate the explicit behaviour of zeros of $h(\lambda)$ so that the Lorentz nature of the  $G$-eigenvectors of $\Omega_0$ can be discerned.  To this end we highlight the following features of the $G$-eigenvectors of $\Omega_0$ -- ascertained from the behaviour of the function  $h(\lambda)$: 
\begin{enumerate}
\item Let us denote the $G$-eigenvector associated with the $G$-eigenvalue $\lambda$ of the matrix $\Omega$ by $X=(a,\, b,\, c,\, d)^T$. By solving the eigenvalue equation $G \Omega_0\, X=\lambda\,X$  i.e., 
\begin{eqnarray}
	\label{eig01}
	&&(n_{0}-\lambda_0)\, a  + n_{1}\, b + n_{2}\, c + n_{3}\, d  =   0, \nonumber \\
	&& n_{1}\, a +  (\alpha_1 +\lambda)\, b  =  0, \nonumber \\ 
	&&	n_{2}\, a + 
	(\alpha_2 +\lambda)\, c  = 0, \nonumber \\  
	&&	n_{3} \, a +  (\alpha_3 +\lambda)\, d=  0. 
\end{eqnarray} 
 
we obtain the explicit form of  the $G$-eigenvector $X$:  
\be
\label{gex}
X=\left(1, \;-\;\frac{n_1}{(\lambda +\alpha_1)}, \;-\;\frac{n_2}{(\lambda +\alpha_2)},
\;-\;\frac{n_3}{(\lambda +\alpha_3)}\right)^T.
\ee
\item  Derivative of the function $h(\lambda)$ (see equation (\ref{hfunction}))  with respect to $\lambda$ at one of its zeros is given by   
\be 
\label{hprime}
h'(\lambda) = -1+\sum_i\, \frac{n_i^2}{(\lambda+\alpha_i)^2}.
\ee 
We identify a striking connection  
\be 
\label{tan}
X^T\, G\, X=-h'(\lambda).
\ee 
between the Minkowski norm of the $G$-eigenvector $X$ (see equation (\ref{gex})) associated with the $G$-eigenvalue $\lambda$ of $\Omega_0$  and the derivative  $h'(\lambda)$. Thus, Minkowski space nature of the $G$-eigenvector  (time-like, null, space-like) of $\Omega_0$ essentially depends on the sign of the derivative $h'(\lambda)$  of the function $h(\lambda)$ at its zeros. 
\item Recall that $h(\lambda)\rightarrow \mp\infty$ as $\lambda\rightarrow\pm \infty$. Furthermore, $h(\lambda)$ has at least two zeros in the region $(\alpha_{\rm max}, \infty)$ (see property 4 of subsection IIIA). Let us denote the largest zero of $h(\lambda)$ (i.e., largest $G$-eigenvalue of $\Omega_0$) by $\lambda_0$.  Since the largest zero $\lambda_0$ appears in the interval $(\alpha_{\rm max}, \infty)$, the slope $h'(\lambda_0)$ of the function  $h(\lambda)$ at  $\lambda_0$ is either negative or zero.  From equation (\ref{tan}) it is seen that the $G$-eigenvector $X_0$ belonging to the largest eigenvalue $\lambda_0$ of $\Omega_0$ obeys  $X_0^T\, G\, X_0\geq 0$. In other words, the $G$-eigenvector $X_0$ associated with the highest $G$-eigenvalue $\lambda_0$ of $\Omega_0$ is either a time-like four-vector (if $h'(\lambda_0)>0$) or null four-vector (if $h'(\lambda_0)=0$). 

It also follows that the largest  $G$-eigenvalue $\lambda_0$ of $\Omega_0$ is  two-fold degenerate when $h'(\lambda_0)=0$, i.e., when $X_0$ is a null four-vector.  

\item  Note that at least one zero must exist between any two discontinuities of $h(\lambda)$.  The tangent $h' (\lambda_r)$ should necessarily be positive for $\lambda_r<\lambda_0$ and hence  it is clear  from (\ref{tan}) that the $G$-eigenvectors $X_r$ of $\Omega_0$  are space-like  four-vectors. 

\item A tetrad $(X_0,\, X_1,\, X_2,\, X_3)$ consisting of one time-like and three space-like $G$-eigenvectors of $\Omega_0$ constitute the columns of a Lorentz matrix~\cite{KNS} $L_{Ic}\in$ SO(3,1) which ensures  the transformation $\Omega_0\rightarrow \Omega_{Ic}=L_{Ic}\, \Omega_0\, L_{Ic}^T={\rm diag}\,(\lambda_0,\, -\lambda_1,\,-\lambda_2,\, -\lambda_3),\ \  \lambda_0\geq\lambda_1\geq \lambda_2\geq \lambda_3\geq 0$, when the $G$-eigenvector $X_0$ is time-like (i.e., $h'(\lambda_0)<0$). 

\item  When  the highest $G$-eigenvalue $\lambda_0$  of $\Omega_0$  is doubly degenerate and the corresponding $G$-eigenvector
is null ($X_0^T\, G\, X_0=0$), one obtains a triad of $G$-eigenvectors, consisting of one null
and two space-like four-vectors. Using this triad it is possible~\cite{AVG1,KNS,supra} to construct a
Lorentz matrix $L_{IIc}$   such that a non-diagonal canonical form~\cite{supra} is realized under the transformation $\Omega_0 \rightarrow \Omega_{IIc}=L_{IIc}\, \Omega_0\, L^T_{IIc}:$
\begin{eqnarray}
	\label{2c}
\Omega_{IIc}&=&\left(\begin{array}{cccc} \chi_0 & 0 & 0 & \chi_0-\lambda_0 \\ 
0 & -\lambda_1 & 0 & 0  \\
0 & 0 & -\lambda_2 & 0  \\
\chi_0-\lambda_1  & 0 & 0 & \chi_0-2\,\lambda_0  
   \end{array}\right),	\nonumber \\ 
\chi_0&=&\left(L_{IIc}\, \Omega_0\, L^T_{IIc}\right)_{00}
\end{eqnarray}
where  $\lambda_0\geq\lambda_1\geq \lambda_2\geq 0$.  It is then realized that positivity of the associated two-qubit density matrix $\rho_{AB}$ demands~\cite{supra} that the $G$-eigenvalues $\lambda_1=\lambda_2$. Therefore the $G$-eigenvalues $\lambda_0$ and $\lambda_1$ are two-fold degenerate in this case.  
\end{enumerate}

In summary, the matrix  $\Omega=\Lambda^T\, G\,\Lambda$ associated with a two-qubit density matrix $\rho_{AB}$  possesses either  
\begin{itemize} 
\item[(i)] a time-like $G$-eigenvector belonging to its largest $G$-eigenvalue $\lambda_0$ and three space-like $G$-eigenvectors 
 \begin{center}
 OR
 \end{center} 
\item[(ii)] a null $G$-eigenvector belonging to  doubly degenerate eigenvalue $\lambda_0$ and two space-like $G$-eigenvectors associated with the remaining  $G$-eigenvalue $\lambda_1=\lambda_2$. 
\item[(iii)]  A tetrad $(X_0,\, X_1,\, X_2,\, X_3)$ consisting of one time-like and three space-like $G$-eigenvectors of $\Omega_0$ form the columns of a Lorentz matrix $L_{Ic}$, which enables a transformation  $\Omega_0\rightarrow \Omega_{Ic}=L_{Ic}\, \Omega\, L_{Ic}^T={\rm diag}\,(\lambda_0,\, -\lambda_1,-\lambda_2,\, -\lambda_3)$ of $\Omega_0$ to a diagonal canonical form. 
\item[(iv)] The traid $(X_0,\, X_1,\, X_2)$ consisting of one null  and two space-like $G$-eigenvectors allows for the construction of a 
Lorentz matrix~\cite{supra} $L_{IIc}$, which transform $\Omega_0$ to its non-diagonal canonical form $\Omega_{IIc}$ given by equation (\ref{2c}), with $\lambda_1=\lambda_1$.
\item[(v)] The $G$-eigenvalues of $\Omega_0$ are non-negative. 
 \end{itemize}

\section{Geometrical visualization of two-qubit states}
Associated with the diagonal canonical form $\Omega_{Ic}$ one obtains the Lorentz canonical form $\Lambda_{Ic}$ for the real parametrization  $\Lambda$ of the two-qubit density matrix $\rho_{AB}$:       
\begin{eqnarray} 
	\label{lambda1c}
	\Lambda\longrightarrow\Lambda_{Ic}&=&{\rm diag}\,  \left(1,\,\sqrt{\frac{\lambda_1}{\lambda_0}},\sqrt{\frac{\lambda_2}{\lambda_0}},\, \pm\, \sqrt{\frac{\lambda_3}{\lambda_0}}\right) 
\end{eqnarray}
where the sign $\pm$ is chosen depending on   ${\rm sign}\left( \det(\Lambda)\right)=\pm$.

Consequently, the two-qubit density matrix $\rho_{AB}$ assumes the following SL(2,C) canonical form (Bell-diagonal): 
\begin{eqnarray}
	\label{rhobd} 	 
\rho^{\,Ic}_{AB}  
		&=&  \frac{1}{4}\, \left( \sigma_0\otimes \sigma_0 + \sum_{i=1,2}\, \sqrt{\frac{\lambda_i}{\lambda_0}}\,\sigma_i\otimes\sigma_i\, 
	 \pm \sqrt{\frac{\lambda_3}{\lambda_0}}\, \sigma_3\otimes\sigma_3 \right).\nonumber \\ 
\end{eqnarray} 

The Lorentz canonical form $\Lambda_{IIc}$ for the real parametrization  $\Lambda$ of two-qubit density matrix $\rho_{AB}$ corresponding to the non-diagonal canonical form of $\Omega_0$ (see equation (\ref{2c})) is given by~\cite{supra}  
\begin{equation} 
	\label{lambda2c}
	\Lambda\longrightarrow\Lambda_{IIc}= \left(\begin{array}{cccc}
		1 & 0  & 0 & 1-s_0 \\ 
		0 & s_1 & 0 & 0 \\ 
		0 & 0 &   -s_1 & 0 \\
		0 & 0 & 0 &  s_0 
	\end{array}\right)
\end{equation}
where  $s_0=\frac{\lambda_0}{\chi_0},\ s_1=\sqrt{\frac{\lambda_1}{\chi_0}}$ (see equation (\ref{2c}) for definition of $\chi_0$).

Correspondingly, one obtains the SL(2,C) canonical form for the two-qubit density matrix:
\begin{eqnarray}
	\label{rho2c} 
	\rho^{\,IIc}_{AB}
	&=&  \frac{1}{4}\, [ \, \sigma_0\otimes \sigma_0 + (1-s_0)\,\sigma_0\otimes \sigma_3   \\ 
	&& \hskip 0.2in + s_1 \,(\sigma_1\otimes\sigma_1 - \sigma_2\otimes\sigma_2) +\,s_0\, \sigma_3\otimes\sigma_3]  \nonumber
\end{eqnarray}

Let us consider projective valued measurements (PVM) on Bob's qubit  by    
\begin{eqnarray}
P&=&\sum_{\mu=0}^{3}\, p_\mu\, \sigma_\mu, P>0,
\end{eqnarray} 
where 
\begin{equation}
\ \ p_0=1, {\mathbf p}=(p_1,p_2,p_3),   \ \ {\mathbf p}^T{\mathbf p}=p_1^2+p_2^2+p_3^2=1. 
\end{equation}
Local PVM on the Bob's qubit  results in collapsed states of Alice's qubit:  
\begin{eqnarray}
\rho_{\mathbf q}&=&\frac{1}{p}{\rm Tr}_B\,\left[\rho_{AB}\,\left(\mathbbm{1}_2\otimes P\right)\right]\nonumber \\ 
&=&\frac{1}{2}\sum_{\mu}\, q_\mu\, \sigma_\mu,\  q_\mu= \frac{1}{p}\,\sum_\nu\,\Lambda_{\mu\,\nu}\,p_\nu
\end{eqnarray}
with probability
\begin{eqnarray*}
	p&=&{\rm Tr}\,\left[\rho_{AB}\,\left(\mathbbm{1}_2\otimes P\right)\right]=\sum_{\mu=0}^{3}\,\Lambda_{0\,\mu}\,p_\mu.
\end{eqnarray*} 

For the Lorentz canonical form $\Lambda_{Ic}$  (see (\ref{lambda1c})) of the two-qubit state $\rho^{\,Ic}_{AB}$ steered Bloch points $\mathbf{q}$ of Alice's qubit  are given by 
\begin{equation}
	\mathbf{q}=\left(q_1=\sqrt{\frac{\lambda_1}{\lambda_0}}\, p_1,\, q_2=\sqrt{\frac{\lambda_2}{\lambda_0}}\, p_2,\, q_3=\sqrt{\frac{\lambda_3}{\lambda_0}}\, p_3\right)
\end{equation} 
which obey  the equation of an ellipsoid   with semi-axes  $(\sqrt{\lambda_1/\lambda_0}, \,\sqrt{\lambda_2/\lambda_0},\, \sqrt{\lambda_3/\lambda_0})$ and center  $(0,0,0)$ inside the Bloch sphere:
\begin{equation}
	\label{eIc}
\frac{\lambda_0\, q_1^2}{\lambda_1}+ \frac{\lambda_0\, q_2^2}{\lambda_2}+ \frac{\lambda_0\, q_3^2}{\lambda_3}=1.
\end{equation}
We refer to this as the {\em canonical steering ellipsoid} of the set of all two-qubit density matrices $\rho_{AB}$ which lie on the SL(2,C) orbit of the Bell-diagonal state $\rho^{\,Ic}_{AB}$ (see (\ref{rhobd})). 

 Corresponding to the second Lorentz canonical form   $\Lambda_{IIc}$ (see (\ref{lambda2c})) one obtains canonical steering spheroid  with its semi-axes lengths $(s_1,s_1, s_0)$  and center $(0,\,0,\, 1-s_0)$: 
 \begin{equation}
 	\label{eIIc}
\frac{q_1^2+q_2^2}{s_1^2}+ \frac{q_3-(1-s_0)}{s_0^2}=1.
 \end{equation}
In other words a shifted spheroid, inscribed within the Bloch sphere, represents  the set of two-qubit states $\rho_{AB}$, which are SL(2,C) equivalent to $\rho_{AB}^{IIc}$ (see  (\ref{rho2c})). 

\section{Illustrative examples} 
In this section we consider specific examples of two-qubit density matrices to  illustrate the properties  of $h(\lambda)$ (see Sec. III) and associated canonical steering ellipsoids. To this end, we begin with examples of real symmetric matrix $\Omega_0$ given in equation (\ref{ac}) and construct the associated function $h(\lambda)$. Plot of $h(\lambda)$ as a function of the continuous variable $\lambda$ depicts discontinuities and roots ($G$-eigenvalues) of the function. Lorentz canonical form of the $4\times 4$ real parametrization matrix $\Lambda$ and the  corresponding SL(2,C) canonical form of the two-qubit density matrix are given.     

\noindent {\bf Example 1:} Consider the real symmetric matrix $\Omega_0$ 
\begin{eqnarray}
\label{arb11}
	\Omega_0&=&\ba{lrrr} 1 & 0.2 & 0.1 & 0.03 \\
	0.2 & -0.3 & 0 & 0 \\
	0.1 & 0 & -0.1 & 0 \\ 
	0.03 & 0 & 0 & -0.5
	\ea
\end{eqnarray}
which clearly has the special form given in equation (\ref{ac}). The $G$-eigenvalues of $\Omega_0$ are obtained by solving the secular equation $\mbox{det}(G\Omega_0-\lambda\,\mathbbm{1}_4)=0$ and  given by $\lambda_0=0.921$, $\lambda_1=0.503$, $\lambda_2=0.366$, $\lambda_3=0.109.$ It is found that the $G$-eigenvector 
$X_0=(0.943, -0.303, -0.115, -0.067)$ belonging to the largest $G$-eigenvalue  $\lambda_0=0.921$ is time-like (because its Minkowski norm is positive:  
$X_0^T\,G\,X_0=0.780$). 

\begin{figure}[h](\ref{arb11})
	\includegraphics*[width=3in,keepaspectratio]{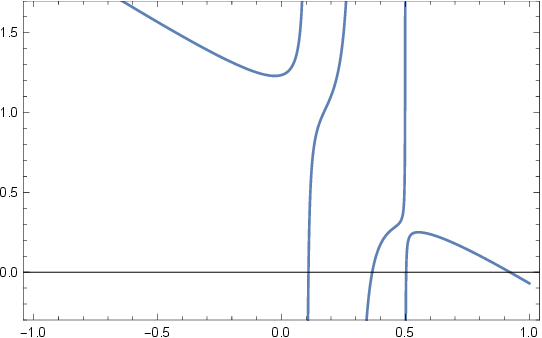}
	\caption{\label{arbone}(Colour online) Plot of $h(\lambda)=1-\lambda-\frac{0.04}{\lambda-0.3}-\frac{0.01}{\lambda-0.1}-\frac{0.009}{\lambda-0.5},$ associated with the real symmetric matrix $\Omega_0$ given by equation (\ref{arb11}) as a function of $\lambda$. At   $\lambda=0.1, \,0.3,\,0.5$ the function $h(\lambda)$ exhibits singularities. Four zeros of  $h(\lambda)$ (which happen to be the $G$-eigenvalues $\Omega_0$ given by equation (\ref{arb11})) occur at $\lambda_0=0.921$,$\lambda_1=0.503$, $\lambda_2=0.366$, $\lambda_3=0.109$. It is seen that the slope  $h'(\lambda_0=0.921)$ is {\em negative}, confirming  that the $G$-eigenvector belonging to the highest $G$-eigenvalue $\lambda_0=0.921$ is time-like.} 
\end{figure}

\begin{figure}[h]
		\includegraphics*[width=3in,keepaspectratio]{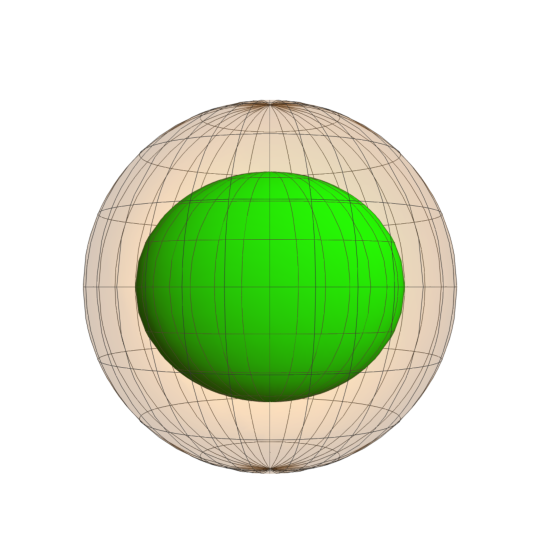}
		\caption{\label{sphd1}(Colour online) Canonical steering ellipsoid representing
			the two-qubit states which are SL(2,C) equivalent to $\rho^{\,Ic}_{AB} $
			of equation (\ref{arb1}). The ellipsoid is centered at the origin of the Bloch sphere with
			semi-axes $\sqrt{\frac{\lambda_1}{\lambda_0}}=0.739$, $\sqrt{\frac{\lambda_2}{\lambda_0}}= 0.630$ and 
		$\sqrt{\frac{\lambda_3}{\lambda_0}}= 0.344$.}
\end{figure}

A plot of  $h(\lambda)=n_0-\lambda-\sum_{i}\, \frac{n_i^2}{\lambda+\alpha_i}$ with $n_0=1$,  $n_1=0.2$, $n_2=0.1$, $n_3=0.03$ and $\alpha_1=-0.3$, $\alpha_{2}=-0.1$, $\alpha_3=-0.5$ associated with $\Omega_0$ in (\ref{arb11})  as a function of the continuous variable $\lambda$ is given in Fig. \ref{arbone}.  It is seen that  $h(\lambda)$ is discontinuous at $\lambda=-\alpha_i,\ i=1,2,3$ i.e., at 0.1,\ 0.3, and 0.5 on the $\lambda$-axis. It is seen that the $G$-eigenvalues of $\Omega_0$  match exactly with the roots $\lambda_0=0.921$,$\lambda_1=0.503$, $\lambda_2=0.366$, $\lambda_3=0.109$ of $h(\lambda)$. Moreover the slope $h'(\lambda_0)$ is negative -- implying that the associated $G$-eigenvector $X_0$ is time-like, in confirmation with the detailed analysis of subsection IIIB. 
Based on this analysis it is seen that the real symmetric matrix $\Omega_0$ of equation (\ref{arb11}) gets transformed to its  canonical form $\Omega_0^{Ic}={\rm diag}\,(0.921,\,0.503,\,0.366,\, 0.109).$  We thus obtain the Lorentz canonical form (see equation (\ref{lambda1c})) of the associated real matrix parametrization of the two-qubit density matrix: 
	$$\Lambda_{Ic}=\mbox{diag}\,\left(1,\,0.739,\,0.630,\,\pm\,0.344 \right).$$
The two-qubit density matrix $\rho^{Ic}_{AB}$ (see equation (\ref{rhobd})) is given by 
\begin{eqnarray}
	\label{arb1} 
	\rho^{\,Ic}_{AB}  
	&=&  \frac{1}{4}\, ( \sigma_0\otimes \sigma_0 + 0.739\, \sigma_1\otimes\sigma_1 \\ 
	&& \ \ + 0.630 \sigma_2\otimes\sigma_2 
	\pm\, 0.344\, \sigma_3\otimes\sigma_3 ). \nonumber
\end{eqnarray} 
The canonical steering ellipsoid representing the set of all  two-qubit states which are on the SL(2,C) orbit of $\rho^{\,Ic}_{AB}$ of equation (\ref{arb1}) is shown in 
Fig.~\ref{sphd1}. 
  
\noindent {\bf Example 2:}
Consider 
\begin{eqnarray}
	\label{arb22} 
	\Omega_0&=&\ba{crrr} 1 & 0.2 & 0.25 & 0 \\
	0.2 & -0.3 & 0 & 0 \\
	0.25 & 0 & -0.15 & 0 \\ 
	0 & 0 & 0 & -0.04
	\ea
\end{eqnarray}
which exhibits the form given by equation (\ref{ac}). We determine the $G$-eigenvalues of $\Omega_0$ (see (\ref{arb22}))  
and obtain
\be 
\label{ex2l} 
 \lambda_0=0.833,\ \lambda_1=0.414,\  \lambda_2=0.202,\  \lambda_3=0.04.
 \ee
 The $G$-eigenvector 
$$X_0=(0.886, -0.332, -0.324, 0)$$ belonging to the largest $G$-eigenvalue  $\lambda_0=0.833$ is found to be time-like i.e., it has positive Minkowski norm $X_0^T\,G\,X_0=0.569$. 

\begin{figure}[h]
		\includegraphics*[width=3in,keepaspectratio]{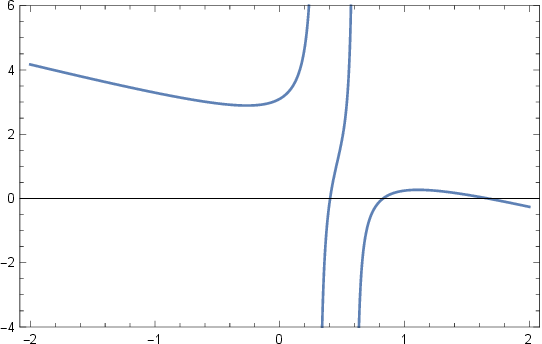}
		\caption{\label{arbtwo}(Colour online) Plot of $h(\lambda)$ given in equation (\ref{hex2}) as a function of  $\lambda$.  It is seen that $h(\lambda)$ exhibits only  {\em two} discontinuities 
		at  $\lambda=0.15,\, 0.3$ and has  {\em three} zeros  at $0.833,\ 0.414,\  0.202$ , which match identically with the $G$-eigenvalues  of 
		$\Omega_0$ of Example 2 (see (\ref{arb22})). Root of the function  $\phi_1(\lambda)=(\lambda-0.04)$ (see case (ii) of Table~1) determines the remaining $G$-eigenvalue is then determined by . It is also seen that the slope of $h'(0.833)$ is negative, confirming that the $G$-eigenvector belonging to the largest $G$-eigenvalues $\lambda_0=0.833$ is {\em time-like} Minkowski four-vector.} 
\end{figure}

A plot of the function  
\be
\label{hex2}
h(\lambda)=0.2-\lambda-\frac{0.04}{\lambda-0.3}-\frac{0.0625}{\lambda-0.15}
\ee  
associated with the matrix  $\Omega_0$ given by (\ref{arb22})  as a function of   $\lambda$ is shown in Fig. \ref{arbtwo}.  Note that  $h(\lambda)$ has {\em two} discontinuities at  0.3 and \ 0.15. Three of the $G$-eigenvalues $\lambda_0,\ \lambda_1,\ \lambda_2$ of $\Omega_0$ (see (\ref{ex2l}))  agree with  the {\em three} zeros $0.202,\, 0.414,\,\ 0.833$ of  $h(\lambda)$. Moreover the slope $h'(0.833)$ is negative -- implying that the associated $G$-eigenvector $X_0$ is time-like. The remaining $G$-eigenvalue $\lambda_3=0.04$ is  equal to the root of the polynomial $\phi_1(\lambda)=(\lambda-0.04)$ (see case (ii) in Table~1).

The Lorentz canonical form of the real matrix parametrization of $\rho_{AB}$ is given by 
\be 
\label{lex2} 
\Lambda_{Ic}=\mbox{diag}\,\left(1,\,0.705,\,0.492,\,\pm\,0.219 \right)
\ee
and the density matrix corresponding to $\Omega_0$ of (\ref{arb22}) is  given by 
 (see equation (\ref{rhobd})) 
\begin{eqnarray}
	\label{arb2} 
	\rho^{\,Ic}_{AB}  
	&=&  \frac{1}{4}\, ( \sigma_0\otimes \sigma_0 + 0.705\, \sigma_1\otimes\sigma_1 \\ 
	&& \ \ + 0.492 \sigma_2\otimes\sigma_2 
	\pm\, 0.219\, \sigma_3\otimes\sigma_3 ). \nonumber
\end{eqnarray}

\begin{figure}[h]
	\includegraphics*[width=3in,keepaspectratio]{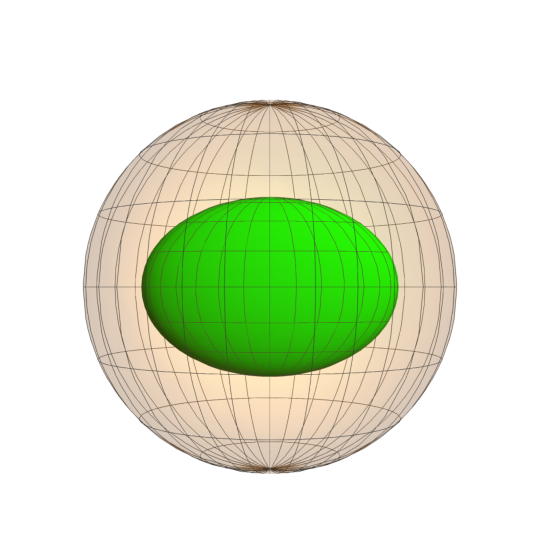}
	\caption{\label{sphd2}(Colour online)  Canonical steering ellipsoid representing the set of all  two-qubit density matrices on the SL(2,C) orbit of     
		$\rho^{I_c}_{AB}$ of equation (\ref{arb2}). The ellipsoid with semi-axes $0.705$, $0.492$, $0.219$ is centered at the origin of Bloch sphere.} 
\end{figure}

The canonical steering ellipsoid representing the set of all  two-qubit states, which are on the SL(2,C) orbit of $\rho^{\,Ic}_{AB}$ of equation (\ref{arb2}) is displayed in Fig.~\ref{sphd2}.

\noindent {\bf Example 3:} 
Let us consider 
\begin{eqnarray}
	\label{2dsym2}
	\Omega_0&=&\ba{lrrr} 0.596 & 0 & 0.148 & 0 \\
	0 & -0.264 & 0 & 0 \\
	0.148 & 0 & -0.183 & 0 \\ 
	0 & 0 & 0 & -0.078
	\ea 
	\end{eqnarray}
for which we find the $G$-eigenvalues as 
\be 
\label{ge3}
\lambda_0=0.533, \, \lambda_1=0.264,\, \lambda_2=0.246,\, \lambda_3=0.078
\ee
and the $G$-eigenvector belonging to the highest $G$-eigenvalue $\lambda_0=0.533$ is given by 
\be 
\label{gev3}
X_0=(0.92, 0, -0.391, 0).  
\ee 
The Minkowski norm $X_0^T\,G\, X_0=0.694$ revealing that the $G$-eigenvector (given in equation (\ref{gev3})) associated with the largest $G$-eigenvalue $\lambda_0=0.533$ of $\Omega_0$ is time-like.   

\begin{figure}[h]
	\includegraphics*[width=3in,keepaspectratio]{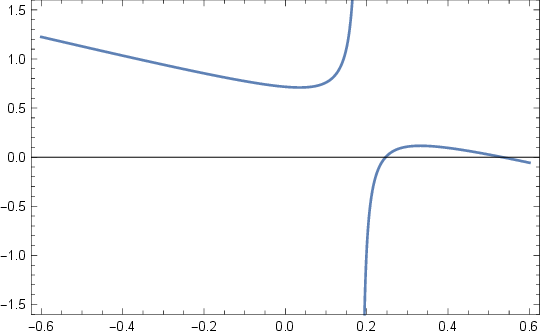}
	\caption{\label{dsn4k2}(Colour online) Plot of  $h(\lambda)=0.596-\lambda-\frac{0.148}{\lambda-0.183}$ associated with $\Omega_0$ given by equation (\ref{2dsym2}). The function $h(\lambda)$ has a single discontinuity at $\lambda=0.183$ and crosses the $\lambda$-axis twice. The  roots of $h(\lambda)$ at $0.246$ and  $0.533$ match with the $G$-eigenvalues $\lambda_1$, $\lambda_0$ respectively (see equation (\ref{ge3})). Two more  $G$-eigenvalues $\lambda_2=0.264, \lambda_3=0.264$ are the roots of the polynomical  $\phi_1(\lambda)=(\lambda-0.264)\,(\lambda-0.078)$ (This corresponds to  case (iii) of Table~1)). It is seen that  $h'(0.533)$ is negative, indicating the {\em time-like} nature of the  the $G$-eigenvector $X_0$ corresponding to the largest $G$-eigenvalues $\lambda_0=0.533$}  
\end{figure}

We have plotted the function $h(\lambda)=0.596-\lambda-\frac{(0.148)^2}{\lambda-0.183}$ corresponding to $\Omega_0$ of equation (\ref{2dsym}) in Fig.~\ref{dsn4k2}. It is seen that $h(\lambda)$ has {\em one} discontinuity at 
$\lambda=0.183$ and {\em two} zeros at $0.246$, $0.533$, which match identically with the $G$-eigenvalues  $\lambda_2$, $\lambda_0$ given in equation (\ref{ge3}). Remaining $G$-eigenvalues are found to be the roots of the polynomial $\phi_1(\lambda)=(\lambda-0.264)\,(\lambda-0.078)$ (see case (iii) of Table~1). The slope $h'(\lambda_0=0.533)$ of $h(\lambda)$ at its largest zero is {\em negative} and ascertains that the $G$-eigenvector belonging to the largest $G$-eigenvalue is {\em time-like}. 

\begin{figure}[h]
		\includegraphics*[width=3in,keepaspectratio]{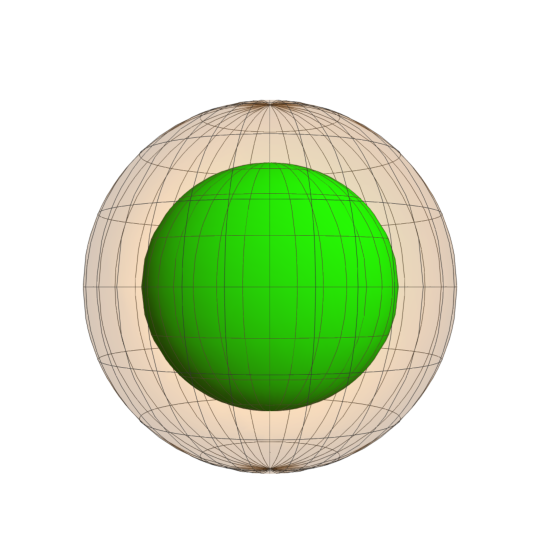}
		\caption{\label{sphd3}(Colour online) Steering ellipsoid representing the set all two-qubit states having SL(2,C) equivalence with $\rho^{\,Ic}_{AB}$ of equation (\ref{2dsym2}). Semi-axes of the ellipsoid  $a_1=0.7$, $a_2=0.68$, $a_3=0.38$.}  
\end{figure}

We obtain the Lorentz canonical form of the real parametrization $\Lambda_{Ic}$ as 
\be
\Lambda_{Ic}=\mbox{diag}\,\left(1,\,0.703,\,0.679,\,\pm 0.382 \right)
\ee
and the corresponding SL(2,C)
canonical form of the two-qubit density matrix is given by 
\begin{eqnarray} 
\label{2dsym2}
	\rho^{\,Ic}_{AB}  
	&=&  \frac{1}{4}\, ( \sigma_0\otimes \sigma_0 + 0.704\, \sigma_1\otimes\sigma_1 \\ 
	&& \ \ + 0.679 \sigma_2\otimes\sigma_2 
	\pm\, 0.384\, \sigma_3\otimes\sigma_3 ). \nonumber
	\end{eqnarray} 
  The canonical steering ellipsoid representing the set of all two-qubit states  connected $\rho^{\,Ic}_{AB}$ of equation (\ref{2dsym2})  via
SL(2,C) transformations is displayed in  Fig.~\ref{sphd3}.

\begin{figure}[h]
		\includegraphics*[width=3in,keepaspectratio]{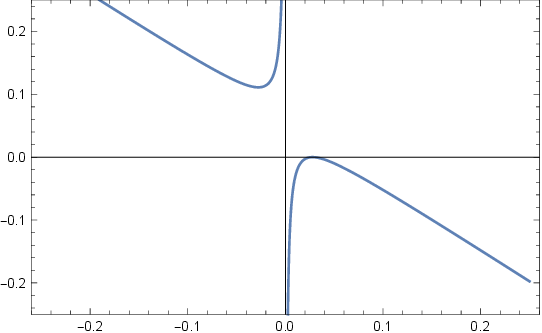}
		\caption{\label{wclass}(Colour online) Plot of $h(\lambda)$ associated with $\Omega_0$ given by equation (\ref{wpi2}). The h-function exhibits only one  discontinuity at $\lambda=0$ and is zero at $\lambda=1/36$.  The slope  $h'(\lambda)$ at $\lambda=1/36$  is zero implying that one of the $G$-eigenvector must be a null four-vector.}
\end{figure}

\noindent {\bf Example 4:} We present an example where  the function $h(\lambda)$ has a single discontinuity and it leads to the SL(2,C) canonical form $\rho^{IIc}_{AB}$ (see equation (\ref{rho2c})) for the two-qubit density matrix. We consider  the real symmetric matrix given by   
\begin{eqnarray}
	\label{wpi2}
	\Omega_0 &=& \frac{1}{36}\ba{crrc} 2 & 0 & 0 & 1 \\
	0 & -1 & 0 & 0 \\
	0 & 0 & -1 & 0 \\ 
	1 & 0 & 0 & 0 
	\ea.
\end{eqnarray} 
We find that the $G$-eigenvalues of $\Omega_0$ are four fold degenerate with $\lambda_0=\lambda_1=\lambda_2=\lambda_3=1/36$. Corresponding $G$-eigenvectors form a triad consisting of one {\em null} and two {\em space-like} four-vectors: 
$$X_0=(1,0,0,-1), X_1=(0,1,0,0), X_2=(0,0,1,0).$$
We have plotted the associated function $$h(\lambda)=\frac{1}{18}-\lambda-\frac{(1/36)^2}{\lambda}$$ in  Fig.~\ref{wclass}. It is evident that $h(\lambda)$ has one discontinuity at $\lambda=0$; a single root at $\lambda=\frac{1}{36}$. The slope $h'(1/36)$ is zero confirming that one of the $G$-eigenvectors is a null four-vector.  

\begin{figure}[h]
	\includegraphics*[width=3in,keepaspectratio]{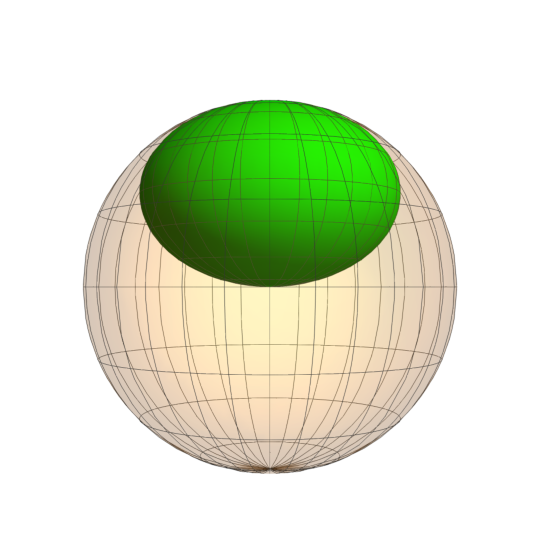}
	\caption{\label{sphdnull}(Colour online) Canonical steering ellipsoid representing the set of all SL(2,C) equivalent two-qubit states $\rho^{IIc}_{AB}$ of equation (\ref{rex4})  is a spheroid centered at $(0,\,0,\,1/2)$ with its semi-axes $\left(\frac{1}{\sqrt{2}},\,\frac{1}{\sqrt{2}},\,\frac{1}{2}\right)$}
\end{figure}

The Lorentz canonical form of the real matrix $\Lambda$  is given by 
\be 
\label{lex4} 
\Lambda_{IIc}=\ba{crrc} 1 & 0 & 0 & \frac{1}{2} \\
0 & \frac{1}{\sqrt{2}} & 0 & 0 \\
0 & 0 & -\frac{1}{\sqrt{2}} & 0 \\ 
0 & 0 & 0 & 0 \ea
\ee
and the associated SL(2,C) canonical form of the two-qubit  density matrix  is  given by (see (\ref{rho2c}))
 \begin{eqnarray}
	\label{rex4} 
	\rho^{\,IIc}_{AB}  
	&=&  \frac{1}{4}\, \left( \sigma_0\otimes \sigma_0 +\frac{1}{2}\sigma_0\otimes\sigma_3  + \frac{1}{\sqrt{2}}\, \sigma_1\otimes\sigma_1  \right. \nonumber \\
	& &\left.  - \frac{1}{\sqrt{2}} \sigma_2\otimes\sigma_2 +\frac{1}{2}\sigma_3\otimes\sigma_3 \right)
\end{eqnarray} 
Geometrical representation of the set of all states which are SL(2,C) equivalent to the two-qubit density matrix given by equation (\ref{rex4}) in terms of a shifted  spheroid with semi-axes $\left(\frac{1}{\sqrt{2}},\,\frac{1}{\sqrt{2}},\,\frac{1}{2}\right)$ and center $(0,\,0,\,1/2)$ is given in Fig.~(\ref{sphdnull})). 

\section{Summary}
We have investigated the two different types of  Lorentz canonical forms for the $4\times 4$ real-matrix parametrization   $\Lambda$ of a two-qubit density matrix $\rho_{AB}=\frac{1}{4}\,\sum_{\mu=0}^4\,\Lambda_{\mu\,\nu}\, \sigma_\mu\otimes\sigma_\nu$. Transformation of two-qubit density matrix $\rho_{AB}$   under local SL(2,C) operations results in Lorentz transformation on the real matrix parametrization $\Lambda$. Based on the observation that $\Lambda$ transforms the set of four-vectors with non-negative Minkowski norm  into itself, it is shown that the spectral properties of the  symmetric matrix $\Omega=\Lambda^T\,G\,\Lambda$,  where $G={\rm diag}\,(1,-1,-1,-1)$ denotes the Minkowski metric, determines the Lorentz canonical forms of $\Lambda$. This leads us to a detailed mathematical analysis, based on Lagrange multiplier approach, to discern the zeros and discontinuities of a function $h(\lambda)$ constructed with the elements of the  matrix $\Omega$. It is shown that the zeros and discontinuities of the function $h(\lambda)$  discern the nature of eigenvalues and eigenvectors of the matrix $G\Omega$. This detailed exploration leads to {\em two} different types of Lorentz canoncial forms $\Lambda_{Ic},\, \Lambda_{IIc}$ for the real matrix parametrization corresponding to the respective SL(2,C) canonical forms $\rho_{AB}^{Ic}$, $\rho_{AB}^{IIc}$ of the two-qubit density matrices. The Lorentz canonical forms $\Lambda_{Ic},\, \Lambda_{IIc}$ enable geometric picturization of the set of all SL(2,C) equivalent density matrices $\rho_{AB}^{Ic}$, $\rho_{AB}^{IIc}$  in terms of  canonical steering ellipsoids inscribed inside the Bloch sphere.

\section{Acknowledgments} 
Sudha, BNK and ARU are supported by Department of
Science and Technology (DST), India, No. DST/ICPS/QUST/
2018/107; HSK is funded by NCN Poland, ChistEra-2023/05/Y/ST2/00005 under the project  Modern Device Independent Cryptography (MoDIC); ASH acknowledges funding from Foundation for Polish
Science (IRAP Project, ICTQT, contract no. MAB/2018/5, cofinanced by EU within Smart Growth Operational Programme).



\end{document}